\documentclass[11pt]{article}
\usepackage{moriond}

\usepackage{amsmath,amssymb,amsthm}

\def\be{\begin{equation}}
\def\ee{\end{equation}}
\def\bea{\begin{eqnarray}}
\def\eea{\end{eqnarray}}

\newcommand{\Photo}{}

\begin{document}
\vspace*{4cm}

\title{Lepton Number Violation with \emph{and without} Majorana Neutrinos \footnote{Talk presented at the 50th Rencontres de Moriond (EW Session), La Thuile, 15 March 2015.}}

\author{Julian Heeck}

\address{Service de Physique Th\'eorique, Universit\'e Libre de Bruxelles, B-1050 Brussels, Belgium}

\maketitle\abstracts{
We discuss the various incarnations of a gauged $B-L$ symmetry: 1) unbroken, it features Dirac neutrinos, neutrinogenesis to create the baryon asymmetry of our Universe, and a potentially light $Z'$ boson; 2) broken by two units, we obtain the standard case of Majorana neutrinos, seesaw and thermal leptogenesis; 3) broken by four units, we find Dirac neutrinos with lepton-number-violating interactions, which can give rise to a new Dirac leptogenesis mechanism. We review and discuss the signatures distinguishing the three scenarios.
}

\section{Introduction}

The observation of a Brout--Englert--Higgs-like scalar boson at the LHC completes the Standard Model (SM). It is however evident that the SM can not be the final description of nature, with 
\begin{itemize}
	\item neutrino masses and mixing, 
  \item dark matter (DM), 
  \item and the baryon asymmetry of the Universe (BAU)
\end{itemize}
among the most pertinent observations that require new physics. All three problems can be solved by relatively simple SM extensions, but there is no unique or (arguable) even simplest solution, so experimental input and theoretical motivation are required to lead the way. In this talk we will consider baryon ($B$) and lepton number ($L$) as guiding principles towards a solution to the three problems above. DM is not the focus here, but we will remark on it parenthetically.

It is well known that the \emph{classical} SM Lagrangian has the accidental global symmetry $U(1)_B\times U(1)_{L_e}\times U(1)_{L_\mu}\times U(1)_{L_\tau}$ due to its particle content/gauge group representations and the requirement for renormalizability. 
Non-perturbative quantum effects -- instantons at zero temperature, sphalerons at $T\neq 0$ -- break both $B$ and $L$ by three units each, so $\Delta (B+L) = 6$, while $B-L$ remains conserved: $\Delta (B-L) =0$. The global symmetry of the \emph{quantum} SM Lagrangian is hence only~\cite{Araki:2012ip}
\begin{equation}
\mathcal{G}_\mathrm{sym} = U(1)_{B-L}\times U(1)_{L_e-L_\mu}\times U(1)_{L_\mu-L_\tau} \,,
\end{equation}
picking a convenient basis in generator space. We know from neutrino oscillations that $U(1)_{L_e-L_\mu}\times U(1)_{L_\mu-L_\tau}$ is a broken symmetry, whereas we have yet to observe a process violating $B-L$. (In fact, no process violating $B$ or $L$ has ever been observed, but we are very confident in the unobservable $\Delta (B+L) = 6$ breaking predicted by theory.)

$\mathcal{G}_\mathrm{sym}$ is the anomaly-free \emph{global} symmetry of the SM Lagrangian at quantum level, and it is tempting to promote it to a \emph{local} symmetry, i.e.~a gauge symmetry alongside $SU(3)_C\times SU(2)_L \times U(1)_Y$. This only requires the introduction of three right-handed neutrinos $\nu_R$, uncharged under the SM gauge group, to cancel anomalies -- a cheap price to pay for such an enlarged gauge group. Furthermore, the quantum numbers of the $\nu_R$ allow us to write down additional couplings
\begin{equation}
\Delta {\mathcal{L}} = -\overline{\nu}_{R} y_\nu H^\dagger L + \text{h.c.} ,
\end{equation}
which automatically give rise to a (Dirac) neutrino mass matrix $m_D = y_\nu \langle H\rangle$ after electroweak symmetry breaking. Promoting the global symmetry of the SM to a local symmetry thus requires neutrino masses \emph{for consistency}, which can be taken as a motivation for this approach.

Flavored subgroups of $\mathcal{G}_\mathrm{sym}$, such as $U(1)_{B+3 (L_e-L_\mu - L_\tau)}$ or $U(1)_{L_\mu-L_\tau}$, make for simple flavor symmetries that can shed light on the leptonic mixing pattern~\cite{Araki:2012ip} and neutrino hierarchies~\cite{Heeck:2011wj,Heeck:2012cd} (see also Ref.~\cite{Heeck:2013bla}). Gauged $L_\mu-L_\tau$ in particular has recently received attention as an explanation for some tantalizing hints in $h\to \mu\tau$~\cite{Heeck:2014qea} and lepton-nonuniversal $B$-meson decays~\cite{Altmannshofer:2014cfa,Crivellin:2015mga,Crivellin:2015lwa} (see contribution by A.~Crivellin in these proceedings).

For simplicity, we will here focus on the unflavored part of $\mathcal{G}_\mathrm{sym}$, i.e.~consider a gauged $B-L$ symmetry. This still requires three right-handed neutrinos, so the argument regarding automatically massive neutrinos from above applies. In the next sections we will explore the different realizations of a gauged $U(1)_{B-L}$ and their different phenomenology, in particular their very different solutions to the problems of neutrinos mass, the BAU, and DM (parenthetically).

\section{Majorana \texorpdfstring{$B-L$}{B-L}}
\label{sec:majorana}

We start with the most popular realization of gauged $U(1)_{B-L}$, in which the symmetry is broken spontaneously by two units, i.e.~$\Delta (B-L) = 2$. For this, a new SM-singlet scalar $\phi_{B-L=2}$ is introduced which carries $B-L  = 2 \ (= -L)$ and can hence couple to the $\nu_R$ via
\begin{equation}
{\mathcal{L}} \ \supset \ -\overline{\nu}_{R} y_\nu H^\dagger L + \tfrac12 \overline{\nu}_R K \nu^c_R \ \phi_{B-L=2}^*+  \text{h.c.} ,
\end{equation}
which gives rise to a right-handed Majorana mass matrix $\mathcal{M}_R = K \langle \phi_{B-L=2}\rangle$ after $B-L$ breaking and ultimately light Majorana neutrino masses via seesaw:
\begin{equation}
\mathcal{M}_\nu \simeq - m_D^T \mathcal{M}_R^{-1} m_D \sim y_\nu^T K^{-1} y_\nu \left( \frac{10^{14}\,\mathrm{GeV}}{\langle \phi_{B-L = 2}\rangle}\right) \,\mathrm{eV} .
\end{equation}
For Yukawa couplings of order one, the $B-L$ breaking scale is untestably high and the only signature of ``Majorana $B-L$'' is neutrinoless double beta decay ($0\nu 2\beta$), mediated by the light Majorana neutrinos. While the $0\nu 2\beta$ rate is definitely non-zero in this scenario, it could still be unobservably small for normal-hierarchy neutrinos if $(\mathcal{M}_\nu)_{ee} \simeq 0$.
Additional signatures arise if the Yukawa couplings are chosen to be small, lowering the right-handed masses below the electroweak scale. In particular, choosing the flavor structure in such a way that one of the right-handed neutrinos, say $\nu_{R,1}$ barely couples to the left-handed neutrinos and has a mass around keV, it can be sufficiently stable to form (warm) DM.\footnote{An alternative approach would be to use the remaining $\mathbb{Z}_2^{L}$ symmetry to stabilize a newly introduced particle with appropriate $B-L$ charge.} The small mixing of $\nu_{R,1}$ then effectively decouples it from the seesaw mechanism, so one of the active neutrinos remains massless.

``Majorana $B-L$'' can also explain the BAU by means of leptogenesis, i.e.~the out-of-equilibrium decay of the heavy right-handed neutrinos $\nu_R\to L H^* , \overline{L} H$ in the early Universe. $CP$-violation arises via loops and results in a lepton asymmetry $\Delta_L$, i.e.~a different number of leptons and antileptons. Since the sphaleron processes ($\Delta B = \Delta L = 3$) are in equilibrium with the rest of the SM plasma at temperatures $T \gtrsim 100\,\mathrm{GeV}$, the lepton asymmetry will partly be converted to a baryon asymmetry $\Delta_B$.

Breaking $B-L$ by two units can hence solve the three main problems of the SM: neutrinos obtain Majorana masses via seesaw, the BAU is explained by leptogenesis, and one can even make one of the right-handed neutrinos stable enough to form DM. This is however not the only viable realization of a gauged $U(1)_{B-L}$, and we will cover two very different scenarios in the next sections.

\section{Unbroken \texorpdfstring{$B-L$}{B-L}}
\label{sec:unbroken}

As already stated in the introduction, we have yet to observe a process that violates $B-L$. It is hence tempting to keep $U(1)_{B-L}$ as an unbroken gauge symmetry~\cite{Heeck:2014zfa}, making $B-L$ a properly conserved quantum number alongside electric charge and color. Neutrinos are then Dirac particles, and one either has to chose the Yukawa couplings very small to obtain the sub-eV required masses, $y_\nu = m_\nu/\langle H\rangle\lesssim 10^{-11}$, or introduce additional new physics that gives a more natural solution.

Surprisingly, even the BAU can be explained in this framework, with a mechanism dubbed neutrinogenesis~\cite{Dick:1999je}. For this, new heavy doublet scalars $\Psi_j$ are introduced which decay out of equilibrium in the early Universe. $CP$ violation via loops can give rise to lepton asymmetries in the decays $\Psi_j\to \overline{L} \nu_R, L\overline{\nu}_R$, which take the form $\Delta_{\nu_L} = - \Delta_{\nu_R} \neq 0$. Lepton number is hence not broken in the decays, but merely distributed among left- and right-handed leptons. The crucial observation is now that the Yukawa couplings $y_\nu = m_\nu/\langle H\rangle$ are \emph{too small} to put the $\nu_R$ in equilibrium with the rest of the SM plasma, and in particular with the sphalerons. These will therefore only see $\Delta_{\nu_L}$, and process it into a baryon asymmetry $\Delta_B$ via the usual $\Delta (B+L ) = 6$ processes, even though the total $B-L$ number of the Universe is zero at all times.

With the BAU and neutrino masses resolved, let us discuss the gauge boson $Z'$ coupled to $B-L$. If massless, the gauge coupling $g'$ is required to be tiny ($g' \lesssim 10^{-24}$) in order to be compatible with tests of the weak equivalence principle. However, since $U(1)_{B-L}$ is abelian, we can actually introduce a $Z'$ mass with the St\"uckelberg mechanism in a gauge-invariant way without breaking the symmetry. This makes the phenomenology of the $Z'$ much more interesting, because the mass is not coupled to neutrino masses, leptogenesis or the weak scale, and can hence sit at any scale. For low masses, constraints in the $M_{Z'}$--$g'$ plane arise from cosmology, astrophysics (stellar evolution), Big Bang nucleosynthesis and colliders~\cite{Heeck:2014zfa}. Unavoidable kinetic mixing results in a $Z'$ coupling to hypercharge and gives rise to additional effects.

As far as DM is concerned, the $Z'$ can be long-lived if the gauge coupling and/or mass are small. The correct abundance can then be obtained by a misalignment mechanism analogous to axions/hidden photons~\cite{Nelson:2011sf}. An alternative way to solve the DM problem in unbroken $B-L$ would be to introduce a new fermion (boson) with even (odd) $B-L$ charge; seeing as all SM fermions (bosons) are odd (even) under $B-L$, the new particle would be stable due to its $U(1)_{B-L}$ charge (similar to the stability of the electron due to the $U(1)_\mathrm{EM}$). A simple freeze-out mechanism using the $Z'$ interactions is then sufficient to obtain the desired DM abundance.

An unbroken gauged $U(1)_{B-L}$ can hence solve all of the three major problems of the SM: neutrinos obtain simple Dirac masses, the BAU can be obtained by neutrinogenesis, and DM can be obtained either with the $Z'$, or with new particles stabilized by the unbroken $U(1)$.

\section{Dirac \texorpdfstring{$B-L$}{B-L}}
\label{sec:dirac}

Let us turn to the third possibility regarding the fate of gauged $U(1)_{B-L}$, where the symmetry is broken -- but not by two units. Breaking $B-L$ by any number $\Delta (B-L)\neq 2$ gives Dirac neutrinos, but since $B-L$ is still broken, this framework still allows for lepton number violation~\cite{Heeck:2013rpa,Heeck:2013vha}.

Seeing as all SM fermions are odd under $B-L$, only $B-L$ breaking by even numbers can be observable (otherwise spin would be violated), so we focus on the simplest $\Delta (B-L)\neq 2$ case: $\Delta (B-L)=4$. Effective operators can be written down without effort:
\begin{align}
 \mathcal{O}^{d=6} &= \overline{\nu}^c_R \nu_R \ \overline{\nu}^c_R \nu_R\,,\label{eq:effective_ops} &
 \mathcal{O}^{d=8}_1 &= |H|^2\ \overline{\nu}^c_R \nu_R \ \overline{\nu}^c_R \nu_R \,, \\
 \mathcal{O}^{d=8}_2 &= (\overline{L}^c \tilde{H}) (H^\dagger L)\ \overline{\nu}^c_R \nu_R \,, &
 \mathcal{O}^{d=8}_3 &= F_Y^{\mu\nu} \overline{\nu}^c_R \sigma_{\mu\nu} \nu_R \ \overline{\nu}^c_R \nu_R \,.
\end{align}
At $d=10$, we only give a selection:
\begin{align}
 \mathcal{O}^{d=10}_1 &= (\overline{L}^c \tilde{H}) (H^\dagger L) \ (\overline{L}^c \tilde{H}) (H^\dagger L) \,, &
 \mathcal{O}^{d=10}_2 &= F_Y^{\mu\nu} (\overline{L}^c \tilde{H}) (H^\dagger L) \  \overline{\nu}^c_R \sigma_{\mu\nu} \nu_R \vphantom{\sum} \,, \\
 \mathcal{O}^{d=10}_3 &= W_a^{\mu\nu}  (\overline{L}^c \tilde{H})  (H^\dagger \tau^a L) \ \overline{\nu}^c_R \sigma_{\mu\nu}  \nu_R \,, &
  \mathcal{O}^{d=10}_4 &= (\overline{u}_R d_R^c) ( \overline{d}_R H^\dagger L) ( \overline{\nu}^c_R \nu_R) \,,
\end{align}
and operators without neutrinos arise at higher dimension still, e.g.
\begin{equation}
  \mathcal{O}^{d=20} = \left[(\overline{(D_\mu L)}^c \tilde{H}) (H^\dagger D_\nu L) \right]^2 \ \supset\ (\overline{e}_L^c W_\mu^+ W_\nu^+ e_L)^2 \,.
\label{eq:charged_leptons}
\end{equation}

Let us present a simple model to show how these effective operators can be obtained and that they are indeed the lowest lepton-number-violating operators, i.e.~$\Delta (B-L) = 2$ processes do not arise. We introduce a scalar $\phi$ with $B-L$ charge $4$ to break the $U(1)_{B-L}$ spontaneously by four units; in order to connect the symmetry breaking to the fermion sector, a second scalar $\chi$ with $B-L$ charge $-2$ is introduced which serves as a mediator and does not acquire a vacuum expectation value (VEV). The important parts of the Lagrangian are
\begin{equation}
- {\mathcal{L}} \ \supset \  \overline{\nu}_{R}   y_\nu  H^\dagger L + \tfrac{1}{2} \overline{\nu}_{R} K \nu_{R}^c \, \chi  + \mu\, \chi^2 \phi + \text{h.c.}
\label{eq:model}
\end{equation}
One can easily realize a scalar potential with minimum at $\langle \chi\rangle =0$, $\langle H\rangle\neq 0 \neq \langle \phi \rangle$, which breaks $SU(2)_L\times U(1)_Y \times U(1)_{B-L}$ to $U(1)_\mathrm{EM}\times \mathbb{Z}_4^L$. An exact $\mathbb{Z}_4^L$ symmetry remains, under which leptons transform as $\ell \to -i \,\ell$ and $\chi \to - \chi$, making the neutrinos Dirac particles but still allowing for $\Delta L = 4$ processes.\footnote{Conservation of lepton number modulo $n > 2$ to forbid Majorana masses was also mentioned in Ref.~\cite{Witten:2000dt}.}
(The remaining $\mathbb{Z}_4$ symmetry could naturally be used as a stabilizing symmetry for a new DM particle, interacting with the SM via the $Z'$ and the scalars.)

\begin{figure}
\centerline{\includegraphics[width=0.85\textwidth]{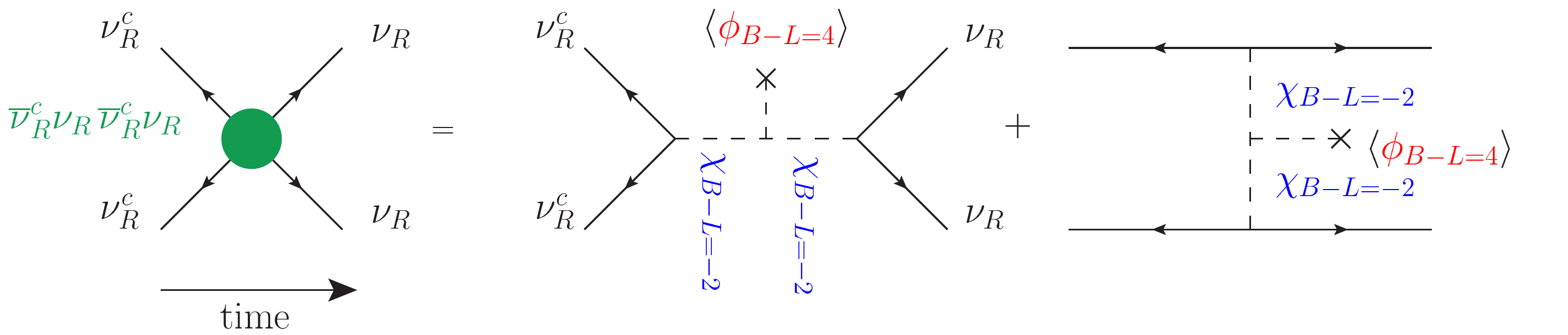}}
\caption{$\Delta (B-L) = 4$ operator $(\overline{\nu}_R^c \nu_R)^2$ realized by exchange of scalars. Arrows show flow of lepton number.}
\label{fig:uv_model}
\end{figure}

Since $\chi$ does not acquire a VEV, the neutrinos will be Dirac particles $\nu = \nu_L + \nu_R$ with mass matrix $m_D = y_\nu \langle H \rangle$, just like in the unbroken $B-L$ case of Sec.~\ref{sec:unbroken}. The VEV of $\phi$ splits the masses of the real and imaginary part of $\chi$ due to the coupling $ \mu \chi^2 \phi$, so we end up with two scalars $\chi_{r,i}$ that couple to $\overline{\nu}_R \nu_R^c$. If these scalars are heavy, we can integrate them out to obtain the $\Delta (B-L) = 4$ operator $(\overline{\nu}_R \nu_R^c)^2$ of Eq.~\eqref{eq:effective_ops} (see Fig.~\ref{fig:uv_model}). Other $\Delta (B-L) = 4$ operators can be obtained by attaching SM interactions, or by going to a left--right extension of this simple model (see below). We stress again that neutrinos are Dirac particles here, and that there are no $\Delta (B-L) = 2$ processes allowed by the symmetry (such as $0\nu 2\beta$).

The $\Delta (B-L) = 4$ interactions can give rise to a new leptogenesis mechanism with Dirac neutrinos that differs qualitatively from the neutrinogenesis mechanism described in Sec.~\ref{sec:unbroken}. For this, we assume several heavy mediator scalars $\chi_j$, which decay out-of-equilibrium in the early Universe. Due to the couplings of Eq.~\eqref{eq:model}, the scalars decay either into $\nu_R\nu_R$ or $\nu_R^c\nu_R^c$, and loop corrections induce a different rate for both channels~\cite{Heeck:2013vha}. After all the scalars have decayed, we thus end up with an asymmetry in the right-handed neutrinos $\Delta_{\nu_R}$. This in itself is not helpful, because the right-handed neutrinos are decoupled from the rest of the SM plasma (which was the main trick in neutrinogenesis). In our case, we \emph{need} them to be in equilibrium, so we have to introduce a second scalar doublet to the SM that has stronger couplings to the $\nu_R$ than the doublet that generates the neutrino mass. Such a model has already been proposed independently of Dirac $B-L$ in order to explain the smallness of Dirac neutrino masses~\cite{Davidson:2009ha}. In this neutrinophilic two-Higgs-doublet model an additional global symmetry ensures that the second doublet couples only to $\overline{L}\nu_R$, and that it only acquires a tiny VEV (say eV). Because of this, the neutrino masses are small even if the Yukawa couplings to the second scalar doublet are large, solving the issue of small Dirac neutrino masses. Even better, the large Yukawa couplings imply that in our leptogenesis scenario the $\Delta_{\nu_R}$ asymmetry is transferred to an asymmetry in the left-handed leptons by the second doublet, and consequently converted to a baryon asymmetry by the sphalerons. As a consequence of the required thermalization of the $\nu_R$, we expect a contribution to the effective number of neutrinos in the early Universe, namely $N_\mathrm{eff} \gtrsim 3.14$, to be tested with future Planck-like experiments.

Above we have seen that $\Delta (B-L) = 4$ processes can be the lowest-order lepton-number-violating effect if neutrinos are Dirac particles, and also that it can lead to a new kind of Dirac leptogenesis mechanism. Compared to the (already hard to measure) $\Delta L = 2$ processes searched for in $0\nu 2\beta$ experiments, it is even harder to probe $\Delta L = 4$ processes directly, because of the high dimensionality of the underlying effective operators. Sensitive nuclear probes analogous to $0\nu 2\beta$ exist -- namely the $0\nu 4\beta$ decay ${}^{150}_{\phantom{1}60}\mathrm{Nd} \to {}^{150}_{\phantom{1}64}\mathrm{Gd} + 4 e^-$ with energy release $Q_{0\nu 4\beta} \simeq 2.08\,\mathrm{MeV}$ testable with existing data from NEMO -- but the expected rates in the toy model from above are unmeasurable small~\cite{Heeck:2013rpa}. It is hence desirable to construct $\Delta (B-L) = 4$ models that can lead to stronger effects, which can be achieved in left--right extensions.

\begin{figure}
\centerline{\includegraphics[width=0.6\textwidth]{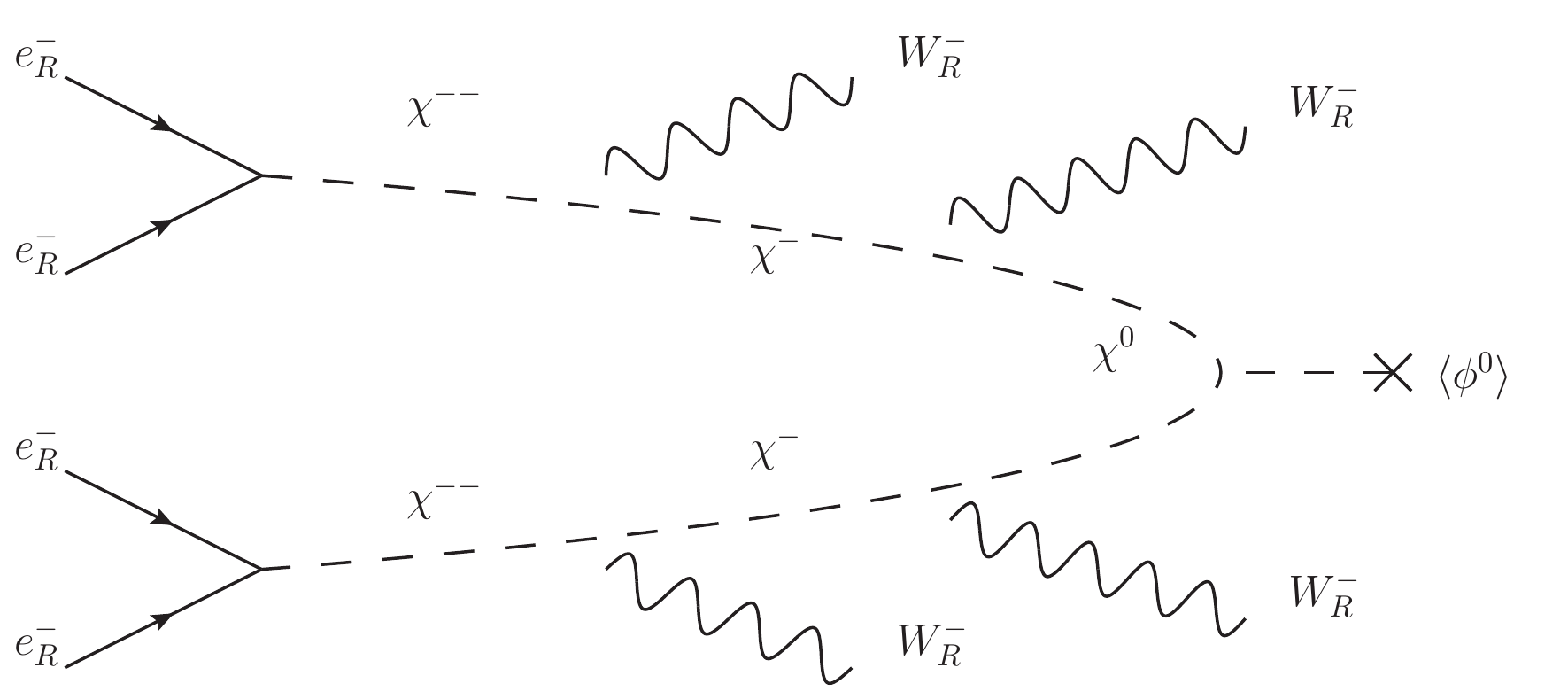}}
\caption{$\Delta (B-L) = 4$ operator $e_R^4 W_R^4$ by exchange of scalars in a left--right model.}
\label{fig:left-right-operator}
\end{figure}

Let us embed the electroweak gauge group $SU(2)_L \times U(1)_Y$ into the left--right symmetry group $SU(2)_L\times SU(2)_R \times U(1)_{B-L}$. Consistency again requires the introduction of right-handed neutrinos (similar to just gauged $B-L$) to complete the right-handed lepton doublet $\Psi_R = (\nu_R, \, e_R)^T \sim (\boldsymbol{1},\boldsymbol{2},-1)$, while the scalar $H$ is promoted to a bi-doublet $H \sim (\boldsymbol{2},\overline{\boldsymbol{2}}, 0)$.
The most common left--right model corresponds to an extension of ``Majorana $B-L$'', i.e.~features Majorana neutrinos. It is however not difficult to also extend ``Dirac $B-L$'' to a left--right model, simply by promoting the scalars $\chi$ and $\phi$ from above to
\begin{eqnarray}
 \chi_R &=&  \frac{1}{\sqrt{2}}
\begin{pmatrix}
 \chi^{-}_R & \chi^{0}_R & 0 \\
 \chi^{--}_R & 0 & \chi^{0}_R \\
 0 & \chi^{--}_R & -\chi^{-}_R 
\end{pmatrix} \sim (\boldsymbol{1},\boldsymbol{3}, -2)\,,  \\
\phi_R &=& \frac{1}{\sqrt{6}} 
\begin{pmatrix}
 \phi^{++}_R & \sqrt{3} \phi^{+++}_R & \sqrt{6} \phi^{++++}_R \\
 \sqrt{3} \phi^{+}_R & -2 \phi^{++}_R & -\sqrt{3} \phi^{+++}_R \\
 \sqrt{6} \phi^{0}_R & -\sqrt{3} \phi^{+}_R & \phi^{++}_R
\end{pmatrix} \sim (\boldsymbol{1},\boldsymbol{5},4)\,.
\end{eqnarray}
The couplings analogous of Eq.~\eqref{eq:model} then take the form (add $\chi_L$ and $\phi_L$ for LR parity)
 \begin{equation}
 {\mathcal{L}} \ \supset  \ y \overline{\Psi}_L H \Psi_R + \kappa  \, \overline{\Psi}_{R} \chi_{R} \Psi_{R}^c  +\mu\, \text{tr}\left[\chi_{R} \phi_{R} \chi_{R}\right] + \text{h.c.} ,
\end{equation}
so $\phi^0_R$ and $\chi^0_R$ play the same role as $\phi$ and $\chi$ from above. Note that the triplet $\chi$ does not acquire a VEV in this model, so the neutrinos are Dirac. $SU(2)_R$ is nevertheless broken above the electroweak scale via $\langle \phi_R\rangle\gg \langle H\rangle$:
\begin{equation}
 M_{W_R^{\pm}}^2 \simeq 2 g_R^2 \langle \phi^0_R\rangle^2 \,, \quad
 M_{Z_R}^2 \simeq 8 (g_R^2 + 4 g_{B-L}^2)  \langle \phi^0_R\rangle^2\,.
\end{equation}
Compared to the toy model from above, it is now possible to consider processes that do not involve neutrinos, and are in particular not suppressed by small neutrino masses (see Fig.~\ref{fig:left-right-operator}). This opens the way towards collider searches for $\Delta L = 4$ processes such as $pp\to 4 \ell^- + 4 W^+$ at the LHC or $e^- e^- \to \ell^+ \ell^+ + 4 W^-$ at a future like-sign lepton collider~\cite{preparation}.

\section{Conclusion}
\label{sec:conclusion}

The incredible success of the Standard Model just deepens the mystery of its anomaly-free global symmetry $U(1)_{B-L}$. Consistently promoting this global symmetry to a local one automatically results in massive neutrinos, amending a major problem of the SM. The matter--antimatter asymmetry of our Universe is also deeply connected to the quantum number $B-L$, and the new particles in the wake of the $U(1)_{B-L}$ are potential candidates for DM.
We presented an overview of the three phenomenologically distinct realizations of a gauged $U(1)_{B-L}$: 1) as an unbroken symmetry with a St\"uckelberg $Z'$, Dirac neutrinos and neutrinogenesis; 2) broken by two units with Majorana neutrinos, seesaw, and leptogenesis; 3) broken by $n\neq 2$ units, e.g.~$n=4$, leading to lepton-number-violating Dirac neutrinos and Dirac leptogenesis.
Experiments will have to decide the fate of $B-L$ and resolve the mystery surrounding it. 

\section*{Acknowledgments}
I thank Werner Rodejohann for collaboration on some of the work presented here, and the organizers of the \emph{Moriond EW 2015}  for financial support and the opportunity to present my results. This work is funded in part by IISN and by Belgian Science Policy (IAP VII/37).

\end{document}